\documentclass[%
 reprint,
 amsmath,amssymb,
 aps,
 prb,
]{revtex4-1}

\usepackage{dcolumn}
\usepackage{bm}
    \usepackage{amsmath,amssymb, mathrsfs}
\usepackage[colorlinks=true,bookmarks=false,citecolor=blue,urlcolor=blue]{hyperref} 

\usepackage{graphicx}
\usepackage{float}
\usepackage{xcolor}
\usepackage{caption}
\begin{document}

\title{Cavity Soliton-Induced Topological Edge States}

\author{Christian R. Leefmans$^{1,\ast}$, Nicolas Englebert$^{2,3,\ast}$, James Williams$^{3}$, Robert M. Gray$^{3}$,\\ Nathan Goldman$^{4}$, Simon-Pierre Gorza$^{2}$, Fran{\c c}ois Leo$^{2}$, and Alireza Marandi$^{1,2,\dagger}$\\
\textit{$^1$Department of Applied Physics, California Institute of Technology, Pasadena, CA 91125, USA.\\
$^2$Service OPERA-Photonique, Universit{\' e} libre de Bruxelles (ULB), Brussels, Belgium.\\
$^3$Department of Electrical Engineering, California Institute of Technology,Pasadena, CA 91125, USA.\\
$^4$CENOLI, Universit{\' e} libre de Bruxelles (ULB), Brussels, Belgium.}\\
$^\ast$These authors contributed equally.\\ $^\dagger$\href{mailto:marandi@caltech.edu}{marandi@caltech.edu}
}

\date{\today}

\maketitle

\textbf{Over the past decade, cavity solitons have attracted substantial attention for their rich dynamics and their myriad potential applications. Recently, there has been growing interest in understanding cavity solitons in systems of coupled resonators, where both new physics and applications can emerge. While numerous works have theoretically studied the interplay between cavity solitons and lattice topology, experimental demonstrations of cavity solitons in topological lattices remain elusive. Here, we experimentally realize cavity solitons in a Su-Schrieffer-Heeger (SSH) lattice and illustrate that the synergy between topology and soliton formation dynamics can induce soliton formation at the boundaries of a topological SSH lattice. Our work illustrates the rich physics of cavity solitons in topological lattices and demonstrates a flexible approach to study solitons in large-scale coupled resonator arrays.}

Since the groundbreaking discovery that photonic systems can support analogs of the quantum Hall effect\cite{raghu_analogs_2008}, photonics has become a medium of choice to study topological phenomena\cite{ozawa_topological_2019}. The role of photonics in topological physics has continued to grow in recent years, thanks to increased interest in the effects of nonlinearity in topological systems\cite{smirnova_nonlinear_2020}. The diverse and controllable sources of nonlinearity available in photonics have enabled exciting fundamental demonstrations, including nonlinear Thouless pumps\cite{jurgensen_quantized_2021}, Floquet topological solitons\cite{mukherjee_observation_2020}, and Laughlin states of light\cite{clark_observation_2020}. Moreover, the combination of nonlinearity and topology may enable the development of robust photonic technologies like topological lasers\cite{zeng_electrically_2020,dikopoltsev_topological_2021} and topological frequency converters\cite{kruk_nonlinear_2019}. Both these fundamental and practical opportunities have fueled interest in understanding the interplay between topology and mode-locked photonic sources, such as dissipative cavity solitons\cite{mittal_topological_2021,fan_topological_2022,tusnin_nonlinear_2023}.

Like topological phenomena, cavity solitons\cite{kippenberg_dissipative_2018} have attracted considerable attention for both their rich dynamics and for their potential to enable transformative technologies (e.g., LIDAR\cite{riemensberger_massively_2020}, optical clocks\cite{newman_architecture_2019}, and on-chip spectrometers\cite{bao_architecture_2021}). While many studies have focused on cavity soliton generation in individual photonic resonators, recently there has been substantial interest in studying cavity solitons in systems of coupled resonators. Preliminary experiments have shown that multi-resonator effects can dramatically influence cavity soliton dynamics, leading to novel phenomena like gear solitons\cite{tikan_emergent_2021,komagata_dissipative_2021}, to practical effects like enhanced conversion efficiencies\cite{xue_super-efficient_2019} and tunable frequency combs\cite{helgason_dissipative_2021}, and to potential applications like soliton-based memories\cite{yuan_soliton_2023}. However, these experiments have been restricted to small system sizes (fewer than 10 resonators), which has limited the dynamics that they can explore and prevented them from studying the interplay between cavity solitons and topological lattices. Therefore, experimentally studying cavity solitons in topological lattices requires developing an alternative approach to realize cavity solitons in large-scale coupled resonator arrays. In turn, such an approach has the potential to accelerate the general study of cavity solitons in coupled resonators.

Time-multiplexed resonator networks have emerged as promising platforms for studying large systems of coupled resonators\cite{leefmans_topological_2022}. By leveraging temporal synthetic dimensions\cite{ozawa_topological_2019-1}, time-multiplexed networks provide a flexible and scalable architecture for implementing a broad variety of topological lattices. In the linear regime, these networks have enabled multidimensional dissipative topological lattices\cite{leefmans_topological_2022}, topological invariant measurements\cite{parto_non-abelian_2023}, and non-Hermitian topological sensing\cite{parto_enhanced_2023}. Moreover, time-multiplexed networks have shown great promise for constructing systems of coupled nonlinear resonators, and they have been adapted to realize both optical Ising machines\cite{marandi_network_2014} and a topological temporally mode-locked laser\cite{leefmans_topological_2022-1}. This broad applicability suggests that time-multiplexed networks might also be adapted to study cavity solitons in temporal topological lattices.

\begin{figure*}
    \centering
    \includegraphics[width=\textwidth]{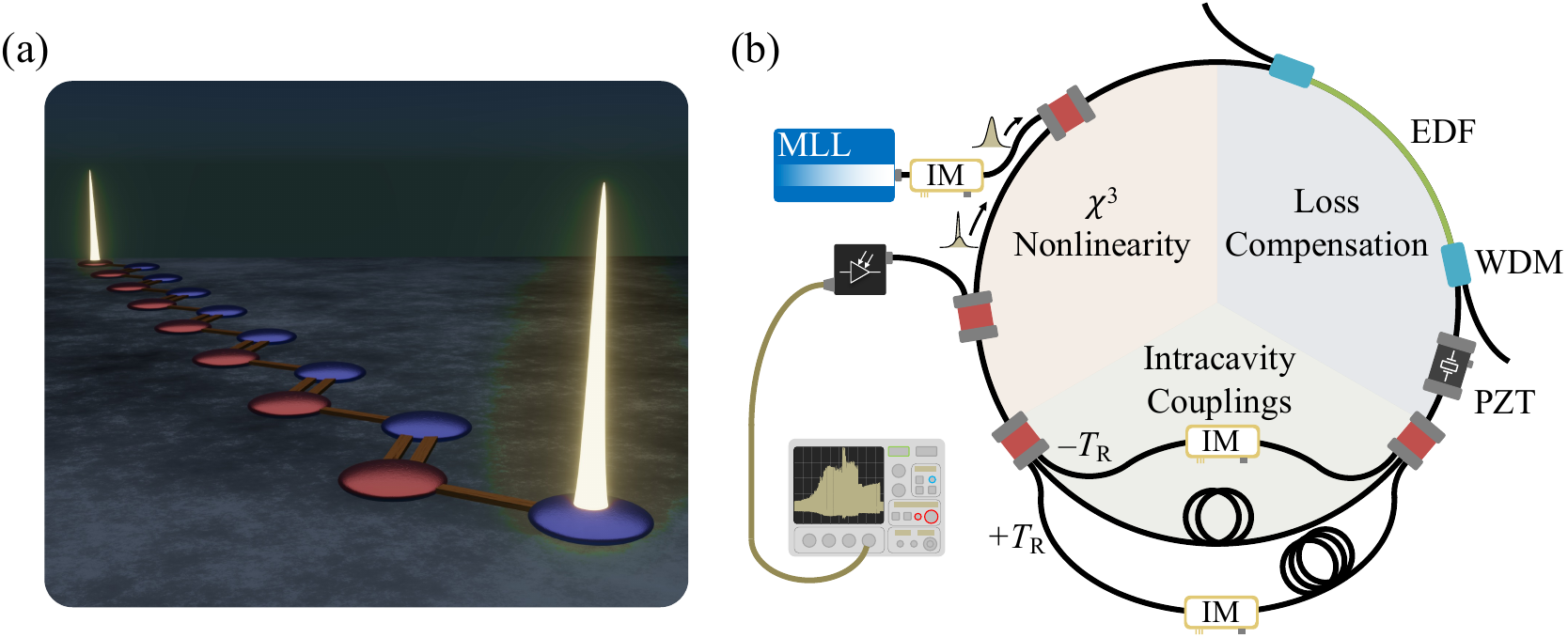}
\caption{{\textbf{Cavity Soliton-Induced Topological Edge States}}~{\textbf{(a)}}~Artistic depiction of cavity soliton-induced topological edge states in a topological Su-Schrieffer-Heeger (SSH) lattice. The SSH lattice consists of a one-dimensional chain with alternating coupling strengths. \textbf{(b)}~Schematic of the time-multiplexed resonator network used to realize cavity soliton-induced topological edge states. This network leverages temporal synthetic dimensions to generate topological lattices capable of supporting cavity solitons.}
    \label{fig:setup}
\end{figure*}

In this work, we use a time-multiplexed resonator network to experimentally study cavity soliton dynamics in temporal topological lattices. Our experiments focus on the Su-Schrieffer-Heeger (SSH) model\cite{su_solitons_1979}, which consists of a dimerized, one-dimensional chain with staggered coupling strengths [see Fig.~\ref{fig:setup}\textcolor{red}{(a)}]. When our system implements a topological SSH lattice, we experimentally demonstrate that the system’s dynamics can favor cavity soliton formation at the boundaries of the topological lattice, leading to what we call cavity soliton-induced topological edge states [Fig.~\ref{fig:setup}\textcolor{red}{(a)}]. We further show that the formation of these edge solitons is robust against the presence of disorder in the topological lattice. Our results reveal an unexplored synergy between topology and cavity soliton dynamics and raise fundamental questions about the relationship between cavity solitons and topological lattices. We conclude by discussing how our system can be adapted to study the interplay between cavity solitons and topology in different types of topological lattices.

To realize cavity soliton-induced topological edge states, we construct the fiber-based time-multiplexed resonator network shown in Fig.~\ref{fig:setup}\textcolor{red}{(b)}. In this network, we use a mode-locked laser to synchronously pump $N$ pulses [Pulse width: $\sim\!5$~ps; Repetition period: $T_{\textrm{R}}\approx8$~ns] in a resonant cavity.  During each roundtrip [Roundtrip time: $T_{\textrm{RT}}\approx544$~ns], a portion of each pulse is siphoned off and divided between two delay lines, whose lengths are carefully chosen to produce nearest-neighbor dissipative couplings\cite{leefmans_topological_2022} between the pulses when the delay lines recombine with the resonant cavity. By letting each pulse in the resonant cavity represent a site in a temporal synthetic lattice, we can map these intracavity couplings to the hopping terms of a tight-binding lattice. Moreover, by using intensity modulators in the delay lines, we can modify these intracavity couplings from pulse-to-pulse, making it is possible to implement couplings that are inhomogeneous, nonreciprocal, and time-varying\cite{leefmans_topological_2022}. In this work, we exploit this flexibility to achieve the staggered couplings of the SSH model.

The intracavity couplings introduced by our network’s delay lines enable us to realize synthetic topological lattices, but generating cavity solitons in our network also requires a double balance of gain and loss and of dispersion and self-phase modulation\cite{kippenberg_dissipative_2018}. While we can use the intrinsic Kerr nonlinearity and group velocity dispersion of our network’s fiber to balance the effects of self-phase modulation and the dispersion, the additional losses introduced by the intracavity couplings and various other elements in the network create challenges in terms of balancing the gain and loss. All together, these losses reduce the cavity finesse (with the delay line couplings turned off) to roughly 1.7. To partially compensate for these losses, we employ an active cavity design\cite{englebert_temporal_2021,englebert_parametrically_2021} by adding a custom erbium-doped fiber amplifier to our resonant cavity [see Fig.~\ref{fig:setup}\textcolor{red}{(b)}]. This gain section increases the finesse of our resonant cavity to up to $\sim\!14.3$, which is sufficient to generate cavity solitons in our system. To verify that we observe solitons in our time-multiplexed network, we present spectrum and autocorrelation measurements in Supplementary Information Sec.~5.

\begin{figure*}
    \centering
    \includegraphics[width=\textwidth]{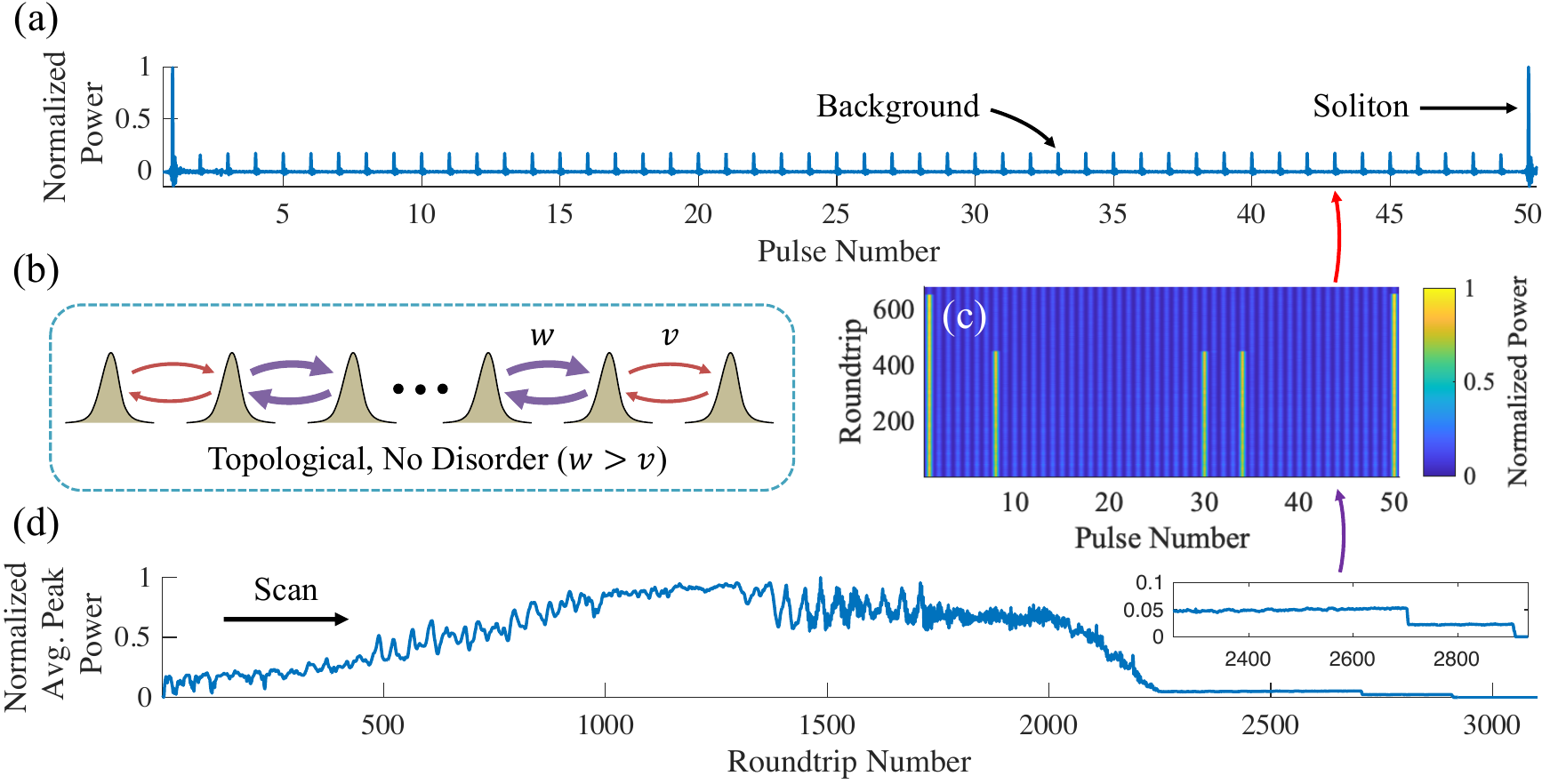}
    \caption{\textbf{Cavity Solitons in a Topological SSH Lattice}~\textbf{(a)}~Dissipative cavity solitons at the boundaries of the topological SSH lattice. This trace is averaged over 150 roundtrips. \textbf{(b)}~Schematic depiction of a topological SSH lattice in the absence of disorder. \textbf{(c)}~Heat map showing the bulk and edge solitons observed in the resonance under study. Note that the pulses are broadened for visibility. \textbf{(d)}~The resonance scan corresponding (a) and (c). The inset shows the soliton steps corresponding to the solitons plotted in (c).}
    \label{fig:topological_no_disorder}
\end{figure*}

To observe cavity soliton-induced topological edge states, we program the delay line intensity modulators in our time-multiplexed resonator network to implement a topological 50-site SSH lattice ($w/v>1$) with a nominal coupling ratio of $w/v=\sqrt{10}$ [see Fig.~\ref{fig:topological_no_disorder}\textcolor{red}{(b)}] and open boundary conditions. We drive all 50 sites of the lattice with our mode-locked laser, and we apply a 60 Hz triangular waveform to the cavity’s fiber phase shifter [labeled “PZT” in Fig.~\ref{fig:setup}\textcolor{red}{(b)}] to sweep the cavity length across the optical resonances. Then, using the procedure detailed in Supplementary Information Sec.~2, we manually tune the length and finesse of our cavity to access an operating regime in which edge solitons appear. An example of these edge solitons is shown in Figs.~\ref{fig:topological_no_disorder}\textcolor{red}{(a,c)}. Here, we clearly see a soliton step in which solitons are present only at the boundaries of the topological SSH lattice.

We demonstrate the consistency with which these edge solitons form by repeating our measurement procedure multiple times across seven different days. Because our measurement procedure involves tuning the length and finesse of the cavity to find the desired operating regime, the finesse of the main cavity can vary between each measurement. However, we estimate that the median main cavity finesse is roughly 6.8, that the median average driving power is roughly 18.4-18.6 mW, and that the larger of the two coupling strengths is on the order of 15\%. For each measurement, we capture resonance scans like that shown in Fig.~\ref{fig:topological_no_disorder}\textcolor{red}{(d)}, and we record whether each resonance in the trace has edge solitons or not. For the topological SSH lattice without additional disorder, we observe 63 resonance scans, 56 of which exhibit solitons on both boundaries, six of which have solitons at only one boundary, and only one which does not have solitons at either boundary. The substantial percentage of resonances that contain solitons on both edges (88.9\%) strongly suggests that the dynamics of the system favors edge soliton formation.

In addition to solitons forming preferentially at the boundaries of the synthetic SSH lattice, we observe that the edge solitons in the disorder-free SSH lattice tend to exhibit larger existence regions than the bulk solitons appearing in the same trace. Indeed, in 43 of the 63 resonance scans (68.3\%) measured in the topological phase of the SSH model, both edge solitons experience a larger existence region than the bulk solitons. In an additional six resonance scans (9.5\%), only one edge soliton experiences a larger existence region than the bulk solitons. We believe that the tendency of the edge solitons to persist at greater detunings is partly the result of our measurement procedure [see Supplementary Information Sec.~2], which enables us to consistently access an operating regime in which we primarily see isolated solitons in the bulk and at the boundaries of our synthetic lattice.  As we shall discuss later, in the topological SSH lattice without disorder, bulk solitons with larger existence regions than the edge solitons tend to occur in ``soliton pairs.''

\begin{figure*}
    \centering
    \includegraphics[width=\textwidth]{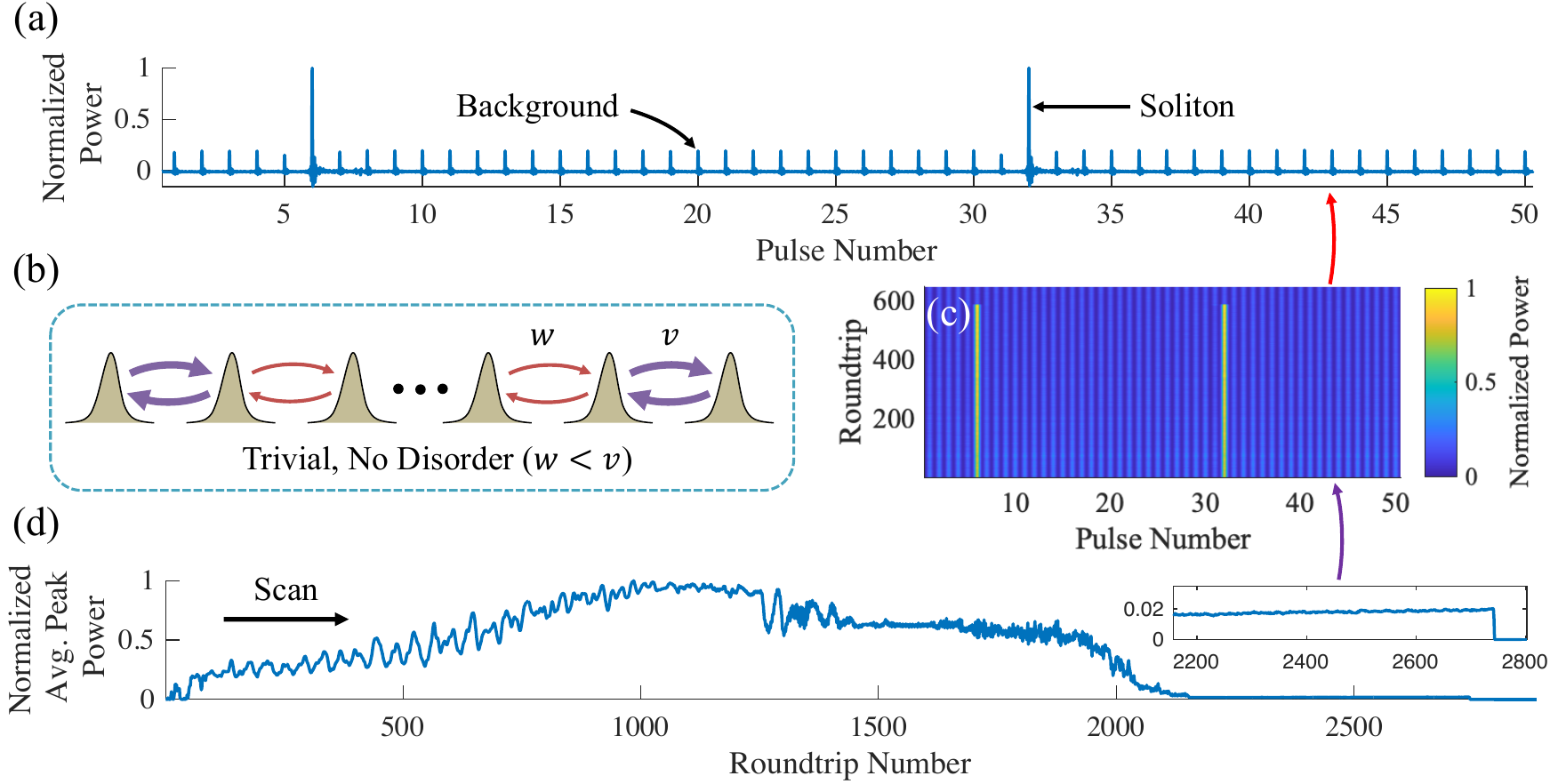}
    \caption{{\textbf{Cavity Solitons in a Trivial SSH Lattice. }}~{\textbf{(a)}}~Trace showing the absence of edge solitons in the trivial phase. This trace is averaged over 150 roundtrips. \textbf{(b)}~Schematic representation of a trivial SSH lattice. \textbf{(c)}~Heat map showing the solitons in the trivial phase. The pulses are broadened for visibility. \textbf{(d)}~Time trace showing the resonance scan corresponding to the solitons presented in (a) and (c). The inset shows the soliton step corresponding to the solitons plotted in (c).}
    \label{fig:trivial_no_disorder}
\end{figure*}

To demonstrate the role of the nontrivial topological lattice in edge soliton formation, we modify the couplings of our time-multiplexed resonator network to implement a trivial SSH lattice ($w/v<1$) with a coupling ratio of $w/v=1/\sqrt{10}$, and we observe multiple resonance scans under similar experimental conditions to those used in the topological phase. In the case of the trivial SSH lattice, we observe 63 resonance scans. One of these resonance scans is presented in Fig.~\ref{fig:trivial_no_disorder}\textcolor{red}{(d)}. Of these scans, no resonances have solitons on both edges, four have a soliton on one edge, 50 have only bulk solitons, and nine exhibit no solitons at all. An example of the bulk solitons captured in the trivial phase is shown in Fig.~\ref{fig:trivial_no_disorder}\textcolor{red}{(a,c)}.  The low rate of edge soliton formation in the trivial phase (6.2\%) suggests that the edge solitons in the topological SSH lattice form due to a synergy between the cavity soliton dynamics and the topologically nontrivial lattice.

\begin{figure*}
    \centering
    \includegraphics[width=\textwidth]{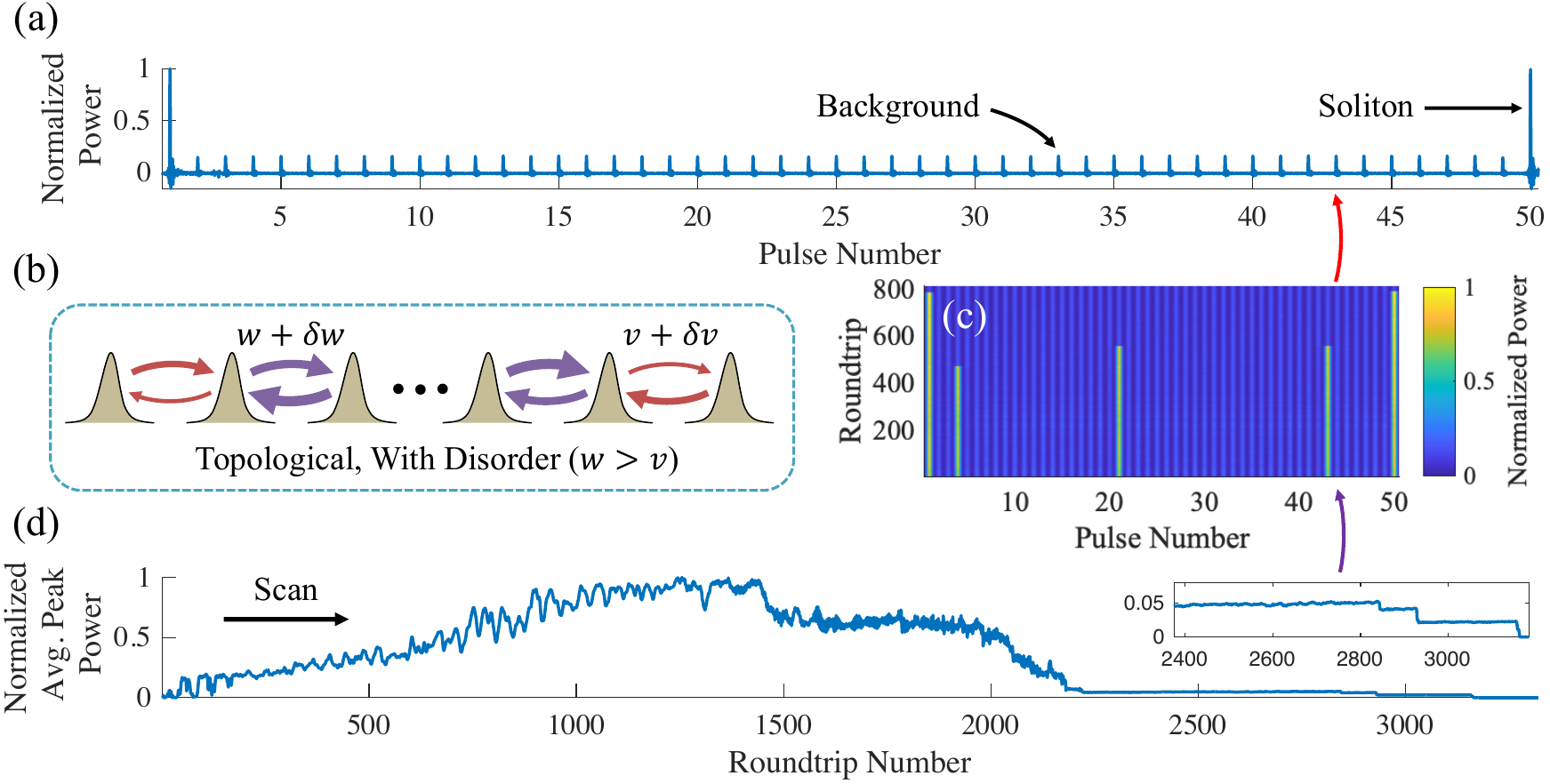}
    \caption{\textbf{Cavity Solitons in a Disordered SSH Lattice} \textbf{(a)}~Trace showing edge solitons in a disordered, topological SSH lattice. This trace is averaged over 150 roundtrips. \textbf{(b)}~Schematic representation of a topological SSH lattice with non-Hermitian coupling disorder. Disorder nominally distributed according to $\operatorname{Unif}(-0.08w,0.08w)$ is added independently to each direction of each coupling. \textbf{(c)}~Heat map showing solitons observed in the disordered SSH lattice. The pulses are broadened for visibility. \textbf{(d)}~Time trace showing a resonance scan corresponding to the solitons shown in (a) and (c). The inset shows the soliton steps corresponding to the solitons plotted in (c).}
    \label{fig:topological_with_disorder}
\end{figure*}

We further highlight the role of the nontrivial topological lattice in edge soliton formation by showing that edge soliton formation is robust against coupling disorder in the topological SSH lattice. Our network implementing a topological SSH lattice with non-Hermitian coupling disorder\cite{leefmans_topological_2022-1} nominally distributed according to $\operatorname{Unif}(-0.08w,0.08w)$ [see Fig.~\ref{fig:topological_with_disorder}\textcolor{red}{(b)}], we measure multiple resonance scans under experimental conditions similar to those used for the measurements above. For the topological SSH lattice with disorder, we observe 54 resonance scans. We find that 51 resonances have solitons on both edges, three have solitons on only one edge, and no resonances are without any edge solitons. As in case of the topological lattice without disorder, the large percentage of resonances that contain solitons on both edges (94.4\%) indicates that the dynamics favors edge soliton formation. We present an example of a resonance scan recorded in the disordered topological phase in Fig.~\ref{fig:topological_with_disorder}.

The above results stand in stark contrast to earlier work on solitons in topological lattices. To the best of our knowledge, the above experiments represent one of the first instances in which cavity solitons have been experimentally demonstrated in topological lattices. Moreover, existing studies of discrete solitons in spatial topological lattices typically operate by selectively exciting soliton solutions\cite{jurgensen_quantized_2021,mukherjee_observation_2020,kirsch_nonlinear_2021,hu_nonlinear_2021,kartashov_observation_2022} in the topological lattice. In stark contrast, the drive in our experiment shows no preference for the edges of our SSH lattices because we excite each site of our synthetic lattice with identical, mode-locked pulses. Instead, the edge solitons in our topological SSH lattices emerge purely due to the complex interplay between the nonlinear soliton formation dynamics and the topologically nontrivial lattice. 

To highlight the role that the soliton formation dynamics play in producing the edge solitons, in Supplementary Information Sec.~6 we simulate resonance scans in linear, dissipatively coupled SSH lattices. While the simulation results for the topological and trivial lattices exhibit some differences, in each case we observe that both the edge and bulk sites in the linear regime remain heavily occupied as the resonance is scanned across. This behavior differs dramatically from that in the soliton regime, where we experimentally witness a clear distinction between the edge behavior in the topological and trivial lattices. Due to the substantial overlap between the edge solitons and the topological edge state of the SSH model, as well as the important role that the soliton formation dynamics have in producing these edge states, we refer to our edge solitons as cavity soliton-induced topological edge states. 

The tendency for solitons to form at the boundary of a topological SSH lattice but not at the boundary of the trivial lattice is believed to occur due to phase and amplitude perturbations introduced by the pulse-to-pulse couplings.  In the modulation instability regime, the pulses in our synthetic SSH lattice experience varying amplitudes and phases, and they perturb their neighboring pulses through the SSH model couplings. In the bulk of the SSH lattice, each pulse is “strongly” coupled to one neighbor and “weakly” coupled to the other. At the edges of the SSH lattice, the pulses are strongly coupled to their nearest neighbors in the trivial phase and weakly coupled to their nearest neighbors in the topological phase. Therefore, both the pulses in the bulk and at the boundary of the trivial lattice appear to experience stronger perturbations relative to the pulses at the boundary of the topological lattice. In the proper operating regime, this appears to create conditions that enable soliton formation at the boundary of the topological lattice while inhibiting soliton formation in the bulk and at the edge of the trivial lattice. When solitons form in both the bulk and at the edges, these perturbations may also explain why isolated bulk solitons tend to annihilate at smaller detunings than the boundary solitons.

The perturbations introduced by the pulse-to-pulse couplings may also explain instances in which we observe bulk solitons with greater existence regions than those of the edge solitons. As we show in Extended Data Fig. 1, in the topological SSH lattice without disorder, the bulk solitons that outlast the edge solitons tend to occur in ``soliton pairs,'' in which the bulk solitons are coupled to one another via the stronger SSH coupling. These solitons appear to constructively interfere with one another, as they each have amplitudes greater than those of the isolated solitons in the system. This constructive interference may help to stabilize the soliton pair against additional perturbations arising from weaker couplings to the solitons’ other neighbors, allowing the soliton pair to exist at greater detunings. Similar arguments may also explain the behavior of soliton pairs at the boundaries of the topological lattice [see Extended Data Fig. 2].

In conclusion, we have introduced and experimentally demonstrated the concept of cavity-soliton induced topological edge states. Using a dissipatively coupled, synthetic SSH lattice, we showed that solitons preferentially form at the boundaries of the lattice in the topological phase but not in the trivial phase. Moreover, we showed that the formation of these edge solitons is robust against disorder. From an experimental perspective, the time-multiplexed network utilized in this work presents distinct opportunities to explore both cavity soliton-induced topological edge states in other topological lattices and to investigate the practical applications of coupled cavity solitons. Moreover, the ability to controllably add different nonlinearities to time-multiplexed networks opens a door to experimentally realize other types of solitons, such as simultons\cite{jankowski_temporal_2018}, in systems of coupled resonators. This capability could contribute to developments in optical gas sensing\cite{gray_cavity-soliton-enhanced_2023}. In future work, we will investigate whether cavity-soliton induced topological edge states can also arise in multidimensional topological lattices.

\section*{Acknowledgments}
The authors acknowledge support from NSF Grants No. 1846273 and 1918549 and AFOSR Award No. FA9550-20-1-0040.  N.E. acknowledges the support of the Belgian American Educational Foundation (BAEF). The authors wish to thank NTT Research for their financial and technical support.

\bibliography{bibliography.bib}
\bibliographystyle{mynaturemag}

\pagebreak

\begin{figure*}
    \centering
    \includegraphics[width=\textwidth]{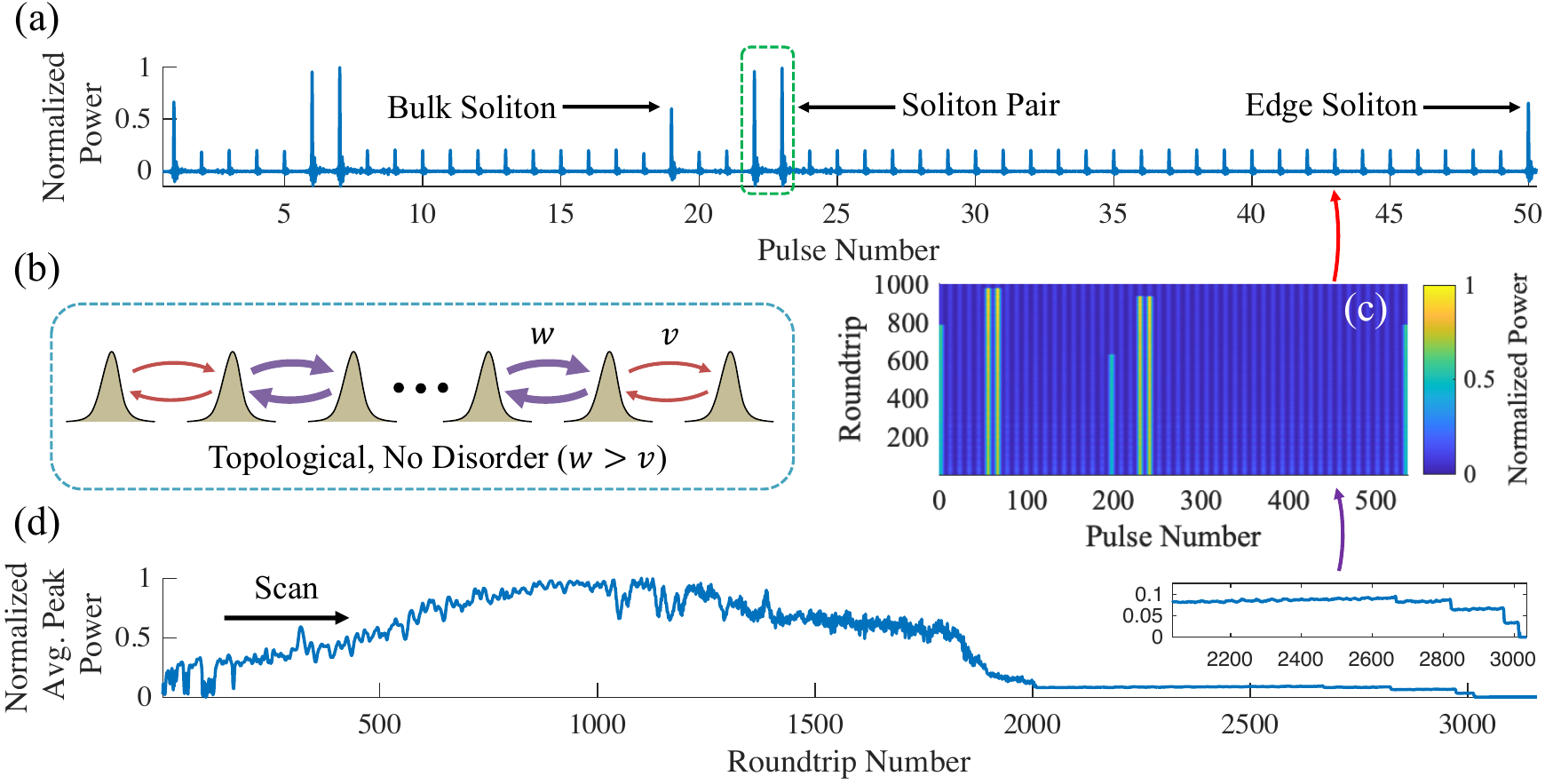}
    \caption*{Extended Data Figure 1: \textbf{Soliton Pairs in the Bulk of an SSH Lattice} \textbf{(a)}~Time trace showing the solitons that appear in the presented resonance scan. We see the emergence of two soliton pairs, which consist of two bulk solitons on neighboring driving pulses. In both soliton pairs, the solitons are coupled by the stronger coupling of the topological SSH lattice. The solitons appear to interfere constructively, as their intensities are larger than those of the isolated solitons in the lattice. \textbf{(b)}~Schematic of a topological SSH lattice without additional coupling disorder. \textbf{(c)}~The soliton pairs exhibit a larger existence region than the isolated solitons in the lattice. \textbf{(d)}~Time trace showing a resonance scan corresponding to the solitons shown in (a) and (c). The inset shows the soliton steps corresponding to the solitons plotted in (c).}
    \label{fig:extended_bulk_soliton_pairs}
\end{figure*}

\begin{figure*}
    \centering
    \includegraphics[width=\textwidth]{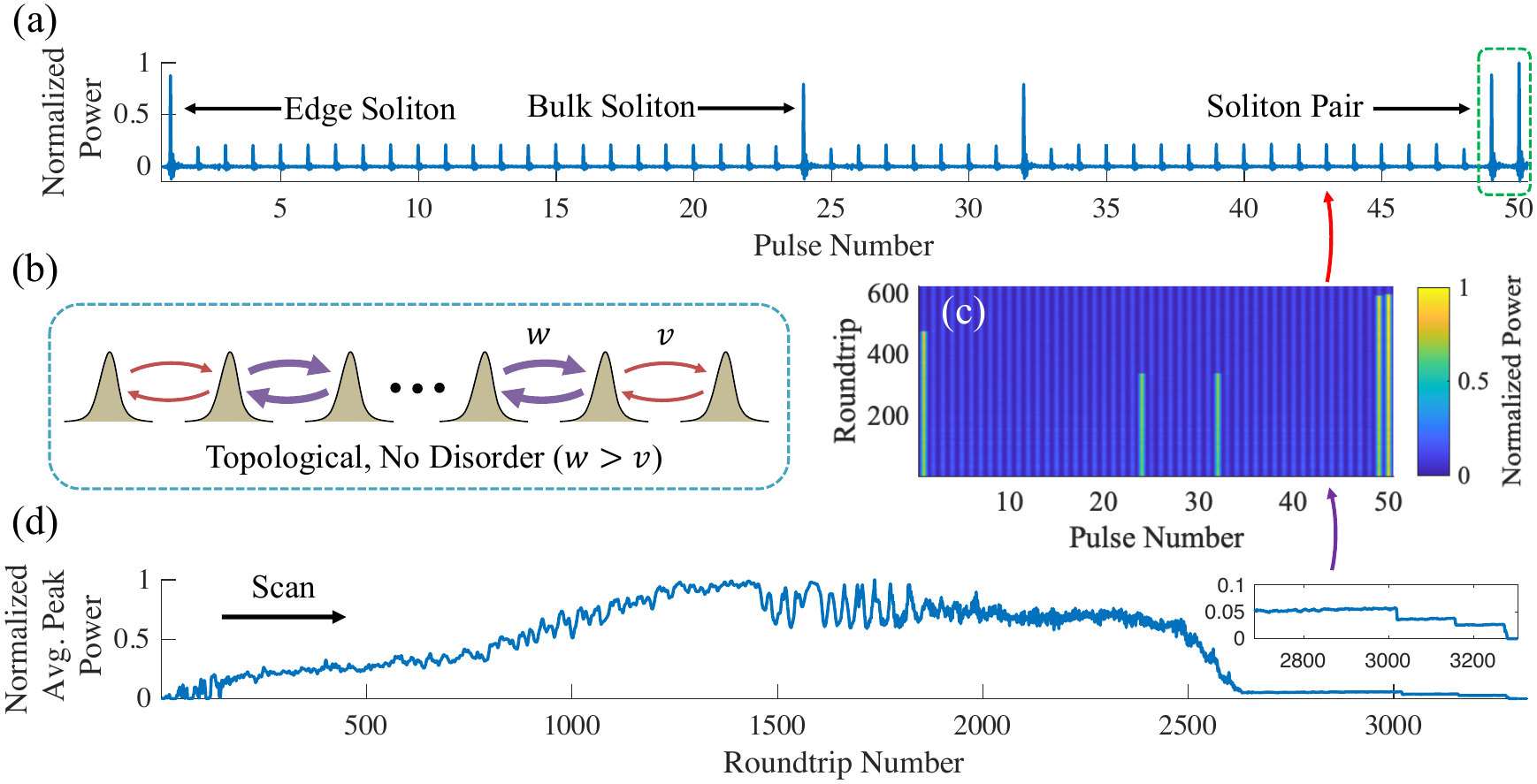}
    \caption*{Extended Data Figure 2: \textbf{Soliton Pairs at the Edge of an SSH Lattice} \textbf{(a)}~Time trace showing the solitons that appear in the presented resonance scan. We see the emergence of a soliton pair consisting of an edge soliton and its neighboring bulk soliton. Here the solitons are coupled by the weaker coupling of the topological SSH lattice. Once again, the solitons appear to interfere constructively because their intensities are larger than those of the isolated solitons in the lattice. \textbf{(b)}~Schematic of a topological SSH lattice without additional coupling disorder. \textbf{(c)}~The soliton pairs exhibit a larger existence region than the isolated solitons in the lattice. \textbf{(d)}~Time trace showing a resonance scan corresponding to the solitons shown in (a) and (c). The inset shows the soliton steps corresponding to the solitons plotted in (c).}
    \label{fig:extended_edge_soliton_pairs}
\end{figure*}

\pagebreak

\end{document}